\documentclass[oneside]{article}
\usepackage{fullpage}
\usepackage[pdftex]{graphicx}
\usepackage{algorithmicx}
\usepackage{algorithm}
\usepackage{algpseudocode} 
\DeclareGraphicsExtensions{.png,.pdf}
\usepackage{hyperref}
\usepackage{verbatim}

\usepackage[small]{caption}
\usepackage[small]{titlesec}
\usepackage{subfig}
\usepackage{amsmath}
\usepackage{amsfonts}
\usepackage{verbatim}
\usepackage{booktabs}
\usepackage[affil-it]{authblk}
\usepackage{listings}
\usepackage{float}
\usepackage{mathtools}
\newcommand{\bs}{\boldsymbol}

\title{Stochastic Volatility Filtering with Intractable Likelihoods}
\author{Emilian Vankov%
  \thanks{Electronic address: \texttt{emilian.r.vankov@rice.edu}; Corresponding author}}
\affil{ Department of Statistics, Rice University}
\author{Katherine B. Ensor%
  \thanks{Electronic address: \texttt{ensor@rice.edu}}}
\affil{ Department of Statistics, Rice University}
\date{\today}
\begin{document}
\maketitle

\begin{abstract}
This paper is concerned with particle filtering for $\alpha$-stable stochastic volatility models. The $\alpha$-stable distribution provides a flexible framework for modeling asymmetry and heavy tails, which is useful when modeling financial returns. An issue with this distributional assumption is the lack of a closed form for the probability density function. To estimate the volatility of financial returns in this setting, we develop a novel auxiliary particle filter. The algorithm we develop can be easily applied to any hidden Markov model for which the likelihood function is intractable or computationally expensive. The approximate target distribution of our auxiliary filter is based on the idea of approximate Bayesian computation (ABC). ABC methods allow for inference on posterior quantities in situations when the likelihood of the underlying model is not available in closed form, but simulating samples from it is possible. The ABC auxiliary particle filter (ABC-APF) that we propose provides not only a good alternative to state estimation in stochastic volatility models, but it also improves on the existing ABC literature. It allows for more flexibility in state estimation while improving on the accuracy through better proposal distributions in cases when the optimal importance density of the filter is unavailable in closed form. We assess the performance of the ABC-APF on a simulated dataset from the $\alpha$-stable stochastic volatility model and compare it to other currently existing ABC filters. 
\end{abstract}

\section{Introduction}
Volatility modeling and estimation has attracted wide interest in the financial community \cite{ensor2013time}. For example, the volatility measure is widely used in portfolio management as well as risk analysis. Since the early 1980's, with introduction of the autoregressive conditional heteroskedastic (ARCH) models \cite{archEng1982} numerous models for quantifying the volatility of asset returns have been developed.  The general idea behind these models is that the conditional volatility at some point in time $t$ depends on its values from previous periods, $t-1, t-2, ..., t-k$, for some $k>0$. Additionally, there have been many extensions to the ARCH discrete time framework, such as GARCH \cite{1986garchBol} and EGARCH \cite{1991egarchNel}.  A goal of these richer models is to capture characteristics of observed volatility in financial data such as volatility clustering, boundedness of volatility, and leverage effects. Volatility clustering is the phenomenon of higher volatility during certain periods of the day. Leverage effects usually refers to the tendency of volatility and returns to be negatively correlated. 
An alternative approach for modeling volatility is the class of stochastic volatility models (SVM) introduced by \cite{1982svmTaylor}, where volatility is generated by a continuous time stochastic process. The discrete representation of SVM is typically given by hidden Markov models (HMM). To make state filtering and parameter estimation more accessible, many authors have assumed that the underlying noise process of the SVM follows a Gaussian distribution. Some empiricists however, have questioned this assumption, due to the presence of asymmetry and heavy tails in stock returns. To account for these characteristics, an alternative solution models the noise process using an $\alpha$-stable distribution. One difficulty in working with this distribution however, is the absence of a functional form of the probability density function.  

We develop an auxiliary particle filter (APF) based on approximate Bayesian computation (ABC) methods \cite{marin2012approximate} for the SVM where the underlying noise is assumed to follow the $\alpha$-stable distribution.  The contribution of our work is two-fold: first, we extend the SVM literature, by proposing a new method for obtaining the filtered volatility estimates. Second, we build upon the current ABC literature by introducing the ABC auxiliary particle filter, which can be easily applied not only to SVM, but to any hidden Markov model. Our paper is organized as follows: Section 2 presents background for hidden Markov models, stochastic volatility models and the $\alpha$-stable distribution. Section 3 sets up the framework of our proposed filter by overviewing approximate Bayesian computation and Sequential Monte Carlo. Section 4 introduces our ABC based auxiliary particle filter (ABC-APF). Section 5 provides simulation results demonstrating the practical implementation of our method to SVM. Finally, we provide some concluding remarks in section 6. 

\section{Background} 
\subsection{Hidden Markov Models}
The evolution of the SVM in discrete time is characterized through the HMM. The HMM in state-space form has an observation and state process describing its evolution. In particular, the discrete time stochastic process is described by an unobserved process $\{X_t\}_{t \geq 0}, \ X_t \in \mathcal{X} \subseteq \mathbb{R}^\mathbb{N}$, called the state process, and another process $\{Y_t\}_{t \geq1 }, \ Y_t \in \mathcal{Y} \subseteq \mathbb{R}^\mathbb{N}$, the observation process. The distribution of the unobserved state is characterized by an initial distribution $X_0 \sim \pi_{\theta}(\cdot)$ and by a transition probability density given by:
\[ X_{t+1} | X_t = x_t  \sim q_{\theta}(\cdot | x_t) \ \ \ \ \ \ \ \    \forall \ t \ \ \text{s.t.} \ \ \ 0 \leq t \leq T\]
for some parameter $\theta \in \Theta \subseteq \mathbb{R}^{\mathbb{N}}$.
Throughout the paper we assume that the observed variables $\{Y_t\}_{t \geq1}$ are conditionally independent given $\{X_t\}_{t \geq0}$, that is,
\[ Y_{t} | X_{0:m} = x_{0:m}  \sim f_{\theta}(\cdot | x_t)   \ \ \ \ \ \ \ \  \forall \ t \ \ \text{s.t.} \ \ \ 1 \leq t \leq m \]
where $X_{i:j} \coloneqq \{X_{i}, X_{i+1}, X_{i+2}, ... \ , X_j\}$.
Generally, one is interested in Bayesian inference based on the posterior densities 
\begin{equation}
p_{\theta}(x_{0:T}|y_{1:T})   \propto p_{\theta}(x_{0:T}, y_{1:T}) = \pi_{\theta}(x_0) \prod_{t=1}^T f_{\theta}(y_t|x_t) q_{\theta}(x_t|x_{t-1})
\label{eq:postden}
\end{equation}
In this case the parameter $\theta$ is assumed to be known and we are interested in obtaining estimates of the states. This set up is the well-known filtering problem \cite{chen2003bayesian}. Alternatively, one can estimate the parameters as well as the filtered states; hence any inference regarding the state-space model is based on the posterior distribution $p(\theta, x_{0:T}|y_{1:T}) = p( x_{0:T}|y_{1:T},\theta)p(\theta) $, 
where $p(\theta)$ denotes the prior distribution.    
Closed forms for the posterior densities above are typically not available for non-linear, non-Gaussian models. Various approximation methods, such as Markov chain Monte Carlo (MCMC)\cite{gelman2013bayesian} and Sequential Monte Carlo (SMC) \cite{doucet2001sequential}, also referred to in the literature as particle filtering, have been developed to sample from the posterior distribution. In our work we focus our attention on SMC methods and providing an approximation to the posterior density for the states with known parameters, namely $p_{\theta}(x_{0:T}|y_{1:T})$ defined by Equation (\ref{eq:postden}). 

\subsection{Stochastic Volatility Models}
 SVMs have many different parameterizations \cite{shephard2009stochastic}. Our starting point is the SVM given by:
\begin{eqnarray}
&&y_t = exp\left({h_t/2}\right)v_t 
\label{eq:ssvmy}\\
&&h_t = \mu + \phi h_{t-1} + \sigma_h w_t 
\label{eq:ssvmx}
\end{eqnarray}
where $y_t$ are observed returns, $h_t$ is latent log-volatility, $\sigma_h$ is the standard deviation of the latent volatility variable, $v_t$ and $w_t$ are noise terms. Different specifications of the noise terms have appeared in the literature, the most common assuming that they are uncorrelated and follow a standard normal distribution \cite{shephard2005stochastic}. This model has been regarded as an approximation to the stochastic volatility diffusion model of \cite{hull1987pricing}. Further consideration of the latent variable in Equation (\ref{eq:ssvmx}) shows that this is an AR(1) process, where $\mu$ is the intercept of the model and $\phi$ is the persistence of the volatility process. To guarantee stationarity of the AR process, we assume that $|\phi | < 1$. To fully specify the model, an initial condition on the latent state variable is necessary. For this model, the initial conditions are given by: $h_0 \sim N\left(\mu / (1-\phi), \ \sigma^2_h / (1 - \phi^2) \right)$. For further discussion of stochastic volatility models we refer the readers to \cite{Taylor:1994aa}and \cite{maddala1996statistical}.

Vast literature exists on state filtering and parameter estimation for this specification of the model, both in frequentist and Bayesian frameworks. The non-linear non-Gaussian model produced by taking the logarithm of the squared returns makes direct application of standard filtering techniques, such as the Kalman filter, incorrect. A possible solution is to approximate $\log (v_t^2)$, which would allow the application of traditional filtering techniques such as the Kalman filter. Even though the distribution itself is not known, \cite{harvey1994multivariate} point out that the mean and variance of  $\log (v_t^2)$ are given to be $1.27$ and $\pi/2$, respectively, and show that an adequate approximation is $\log(v_t^2) \sim N(1.27, \pi^2/2)$. Once the observation equation is transformed to a linear Gaussian form, the Kalman filter can be applied to obtain estimates of the volatility. In particular, the authors propose a quasi-maximum likelihood technique for parameter estimation.  

In the Bayesian framework, the work of \cite{gordon1993novel} was the first to introduce particle filtering for non-linear, non-Gaussian state estimation. Since this seminal work, there has been much literature devoted to improving the particle filter, as well as its applications to hidden Markov models. In particular, \cite{pitt1999filtering} developed the auxiliary particle filter (APF) and applied it to the nonlinear stochastic volatility model discussed above. For a review on particle filters and their applications to financial models, we refer the reader to \cite{creal2012survey}. Bayesian estimation of parameters and the filtering of states has been carried out via MCMC. The advantages of these powerful methods were first noted in \cite{jacquier2002Bayesian}, where standard MCMC chains were employed to obtain draws from the posterior distribution of all parameters as well as the log-volatility. Later, \cite{kim1998stochastic} developed a block sampling MCMC scheme for the model in order to improve the efficiency of the algorithm.  

While the model discussed so far provides a simple and elegant framework for studying volatility, the assumption of normality in the returns is not realistic, as heavy-tails and skewness are typical features of financial returns (\cite{mandelbrot1967variation}, \cite{fama1965behavior}). To account for the heavy-tails observed in returns, the majority of existing literature has focused on extensions of SVM in which the noise term of the observation equation is modeled by a t-distribution. Examples of such works include \cite{harvey1994multivariate} and more recently \cite{chib2002markov}. 

An attractive flexible parametric distribution, the $\alpha$-stable distribution, has the ability to capture the heavy-tailed behavior observed in the returns while allowing for skewness in the distribution. A problem that the use of $\alpha$-stable distribution poses, however, is the inability to express the probability density function in closed form. Therefore, traditional filtering techniques, including inference based on maximum likelihood and direct application of MCMC methods, are unavailable. This has led to the development of alternative methods for parameter estimation and likelihood inference. For example, \cite{mcculloch1986simple} proposed an estimation based procedure that relies on quantiles, whereas \cite{press1972estimation} propose a method of moments approach. In the simpler case of a symmetric $\alpha$-stable distribution,  \cite{nikias1995signal} propose a parameter estimation method based on negative and fractional moments.  The recent development of fast computing has allowed likelihood-based inference based on approximation of the unknown density by fast Fourier transform methods (FFT) of the characteristic function \cite{menn2006calibrated}. In the Bayesian framework, the posterior estimation is carried out by an auxiliary Gibbs sampler  \cite{buckle1995Bayesian}  integration inversion of the characteristic function within MCMC \cite{Lombardi2007alphaMCMC}. We propose a method for obtaining the filtered latent volatility from an SVM with errors generated from a stable distribution, with known parameters, based on ABC. 

\subsection{Stable Distributions}
The stable distribution, denoted here by $SD(\alpha, \beta, \delta, \gamma)$, is a parametric family of distributions characterized by four parameters, the stability parameter $\alpha \in (0,2]$, the skewness parameter $\beta \in [-1,1]$, the scale parameter $\gamma \in [0, \infty)$ and the location parameter $\delta \in (- \infty, \infty)$. The $\alpha$-stable distribution is typically described by its characteristic function. There are several different parameterizations that have appeared in the literature, each leading to different forms of the characteristic function. In our case we represent the characteristic function as:
\begin{equation}
\psi(t) =  \mathbb{E}\left[\exp\{itX\} \right] = 
\begin{cases}
\exp\left\{-\gamma^{\alpha}|t|^{\alpha} \left[1 -i\beta(\tan \frac{\pi\alpha}{2})(sign(t)) \right] +i\delta t\right\}, & \text{if }\alpha \neq 1\\
\exp\left\{-\gamma|t| \left[1 + i\beta\frac{2}{\pi}(sign(t))\log(t) \right] +i\delta t\right\}, & \text{if }\alpha = 1
\end{cases}
\label{eq:chfalpha}
\end{equation}
The pdf of the $\alpha$-stable distribution is consequently given by: 
\[ f(x) = \frac{1}{2\pi} \int_{\mathbb{R}} \psi(t)e^{-ixt}dt\]
The characteristic function in equation (\ref{eq:chfalpha}) does not yield a closed form solution for the pdf \cite{nolan1997numerical}. Some special cases of the stable distribution are the normal distribution when $\alpha = 2$, the Cauchy which has $\alpha = 1$, $\beta = 0$, and the Levy distribution where $\alpha = 1/2$ and $\beta = 1$. For a more in-depth overview of the properties and the theory of stable distributions we refer the reader to \cite{nolan1997numerical} and \cite{zolotarev1986one}.

To conclude this section, we introduce a likelihood-free method for simulating random variables from the stable distribution (see \cite{chambers1976method}). Being able to simulate from this distribution without knowing the likelihood is necessary for our proposed ABC based filter. 

Let $U \sim$ Uniform$(-\frac{\pi}{2}, \frac{\pi}{2})$ and $W \sim$ Expo(1). Define $\psi =\arctan(\beta \tan(\pi \alpha/2))$ for all $\alpha \neq 1$  For any $\beta \in [-1,1]$ the random variable 
\begin{equation}
X = 
\begin{cases}
\frac{\sin \ \alpha (\psi + U)}{(\cos \ \alpha \psi \ \cos \ U)^{1 / \alpha}}\left[\frac{\cos(\alpha \psi + (\alpha -1)U)}{W} \right]^{(1-\alpha)/ \alpha}, & \text{if }\alpha \neq 1\\
\frac{2}{\pi}\left[(\frac{\pi}{2} + \beta U)\tan \ U - \beta \log(\frac{\frac{\pi}{2}W \cos \ U}{\frac{\pi}{2}+\beta W})  \right],
& \text{if }\alpha = 1
\end{cases}
\end{equation}
is distributed as $\mathcal{SD}(\alpha, \beta,1,0)$. One can obtain a sample from a $\mathcal{SD}(\alpha, \beta,\gamma,\delta)$ by applying the following transformation to the random variable X as defined above:
\begin{equation}
Y = 
\begin{cases}
\gamma X + \delta, & \text{if }\alpha \neq 1\\
\gamma X + (\delta + \beta\frac{2}{\pi}\gamma \ log \ \gamma),
& \text{if }\alpha = 1
\end{cases}
\end{equation}

\section{Sequential Monte Carlo and Approximate Bayesian Computation}
\subsection{ABC for posterior estimation}
With recent decreases in computational cost, Bayesian statistical methods have become more attractive. In particular, the development of  MCMC methods has enabled practitioners to model and analyze complex systems in many different fields. Generally, one wants to estimate some parameter $\bs{\theta} = (\theta_1, \theta_2, ... , \theta_d) \in \Theta^d$ for a given model $f(\cdot | \bs{\theta})$. In Bayesian framework, $\bs{\theta}$ is assumed to have some prior distribution $\pi(\bs{\theta})$. Any inference about the posterior distribution $p(\bs{\theta}|\bs{x})$ is based on 
\[ p(\bs{\theta}|\bs{x}) = \frac{f(\bs{x}|\bs{\theta})\pi(\bs{\theta})}{\int_\Theta f(\bs{x}|\bs{\theta})\pi(\bs\theta)d\bs{\theta}}  \] 
where $\bs{x} = (x_1, x_2, ...  , x_n) \in \mathcal{X} \subseteq \mathbb{R}^n$ is the observed data set generated from the model $f(\cdot | \theta)$. In most situations the posterior distribution is not available in closed form and MCMC methods of sampling, such as Metropolis-Hastings \cite{hastings1970monte}, are required. Although, the Markov chains generated from such sampling methods converge to the posterior distribution \cite{gelman2013bayesian}, successful implementation depends on the knowledge of the likelihood function $f(\bs{x}|\bs{\theta})$. Additionally if the likelihood is very complicated then the Markov chain might take a very long time to converge to its stationary distribution. 

In contrast, ABC methods rely on making inferences about a parameter $\bs{\theta}$ via an approximate posterior distribution given by:
\begin{equation}
p(\bs{x}_s, \bs{\theta}| \bs{x}) = \frac{K_{\epsilon}(\psi(\bs{x|\theta}) - \psi(\bs{x}_s|\bs{\theta}))f(\bs{x}_s| \bs{\theta}) \pi(\bs{\theta})}{\int_{\Theta} K_{\epsilon}(\psi(\bs{x|\theta}) - \psi(\bs{x}_s|\bs{\theta}))f(\bs{x}_s| \bs{\theta}) \pi(\bs{\theta})d\bs{\theta} 
}
\label{eq:one}
\end{equation}
where $\textbf{x}_s$ is a simulated dataset from the model $f(\cdot | \bs{\theta})$, $\psi$ is a summary statistic of the data such that, $\psi:\mathcal{X} \rightarrow \Psi$, where $\Psi \subseteq \mathbb{R}^m$ and $ m \leq n$; $K_{\epsilon}(\bs{y}) = \epsilon^{-1} \ K(\bs{y}/\epsilon)$, where $\epsilon > 0$ is the bandwidth and $K$ satisfies $\int K(\bs{y}) d\bs{y} = 1$. Here we have defined $K$ as a general kernel. Most existing research relies on using uniform kernel. However, as discussed in \cite{peters2012sequential} it has not been shown that the use of this kernel is optimal in any sense. In our work, we propose to use the normal kernel instead. While ABC methods relax the assumption of knowing the likelihood of the model exactly it assumes that one can simulate data from the model. As the value of $\epsilon$ plays an important role in the performance of ABC algorithms, care has to be taken in its choice. In particular, smaller values of $\epsilon$ lead to more accurate results but increase the computational burden of the algorithm. One can show that as $\epsilon \rightarrow 0$ and if the summary statistics $\psi$ are sufficient, the true posterior distribution $p(\bs{\theta} | \bs{x})$ is recovered \cite{peters2012sequential}. An adaptive ABC-SMC is developed by \cite{del2012adaptive}, who update the value of $\epsilon$ based on the proportion of "alive" particles at each iteration of the algorithm.
Due to their simplicity and ability to provide estimates from the approximate posterior distribution when the likelihood of the model is unknown, ABC methods have become very popular. \cite{2003MTabc} introduce a MCMC algorithm in ABC framework and show that the resulting Markov chain admits the desired posterior distribution as the stationary distribution. In particular, the Metropolis-Hastings algorithm is modified to replace the evaluation of the likelihood with an ABC data simulation step. More recently, as SMC methods have been becoming more popular in the literature, ABC counterparts for those algorithms have been developed \cite{2007STabc}, \cite{toni2009approximate},  \cite{beaumont2008adaptivity}. For further details on ABC we refer the reader to \cite{marin2012approximate}. In the next section with turn our attention to Sequential Monte Carlo methods and introduce our ABC auxiliary particle filter.

\subsection{SMC and the Auxiliary Particle Filter}
In Sequential Monte Carlo methods one is interested in posterior inference for $p(x_{0:t}| y_{1:t})$. Using the definition of conditional distribution and some algebra one has:
\[p(x_{0:t}| y_{1:t}) = \frac{p(x_{0:t}, y_{1:t})}{p(y_{1:t})} = p(x_{0:t-1}|y_{1:t-1})\frac{q(x_t|x_{t-1})f(y_t|x_t)}{p(y_t | y_{1:t-1})} \]
Note in the above that $p(x_{0:t-1}|y_{1:t-1})$ is the distribution of interest at time $t-1$, and $f$ and $q$ are the observation and state densities respectively, as defined in section 2. If we assume that at $t-1$ we have a collection of random samples $\{\hat{X}_{0:t-1}^{(i)}, i = 1, ..., N\} \sim \hat{p}(x_{0:t-1}|y_{1:t-1})$. Then we can construct the empirical measure
\[\hat{p}^N(x_{0:t-1} | y_{1:t-1}) = \frac{1}{N} \sum_{i=1}^N w_{0:t-1}^{(i)}\delta_{\hat{x}^{(i)}_{0:t-1}}\]
In the above $w_{0:t-1}^{(i)}$ are the weights associated with the realizations of the particles $\{\hat{X}_{0:t-1}^{(i)}, i = 1, ..., N\}$. The dirac measure, $\delta$ is given by:
\[
\delta_x (A) = \left\{
        \begin{array}{ll}
            1 & \mbox{if}  \ \ x \in A \\
            0 & \mbox{if} \ \ x \notin A
        \end{array}
    \right.
\] 
Our goal is to obtain $\hat{p}^N(x_{0:t} | y_{1:t})$ an approximation to $p(x_{0:t} | y_{1:t})$. One way to obtain the approximate distribution at $t$ is to sample $\{\tilde{X}_{t}^{(i)}, i = 1, ..., N\}$ and obtain
\[\{(\hat{X}_{0:t-1}^{(i)}, \tilde{X}_{t}^{(i)}), i = 1, ..., N\}  \sim g(x_{0:t}|y_{1:t}) \]
However, $g(x_{0:t}|y_{1:t})$ is not the target distribution. We can use importance sampling to account for the discrepancy between the two distributions by weighing the samples. That is we construct the weights as follows:
\begin{eqnarray}
w_t^{(i)} &\propto& \frac{p(x_{0:t}^{(i)}|y_{1:t})}{g(x_{0:t}^{(i)}|y_{1:t})} \nonumber \\
&\propto& \frac{p(x_t^{(i)}, y_t |x_{0:t-1}^{(i)}, y_{1:t-1})p(x_{0:t-1}^{(i)}|y_{1:t-1}) }{g(x_t^{(i)} |x_{0:t-1}^{(i)}, y_{1:t})g(x_{0:t-1}^{(i)}|y_{1:t-1})   } \nonumber \\
\nonumber \\
&\propto& \frac{f(y_t | x_t^{(i)})q(x_t^{(i)} | x_{t-1}^{(i)})}{g(x_t^{(i)} |x_{0:t-1}^{(i)}, y_{1:t})} \ w_{t-1}^{(i)} 
\label{eq: wupdate}
\end{eqnarray}
In the last expression above $w_{t-1}^{(i)}$ are the weights from the previous step. After normalizing the weights we can construct the empirical measure
 \[\hat{p}^N(x_{0:t} | y_{1:t}) = \frac{1}{N} \sum_{i=1}^N w_t^{(i)}\delta_{\hat{x}^{(i)}_{0:t}} \approx p(x_{0:t} | y_{1:t})\] 
There are many guidelines on how to select the importance density function $g(x_t |x_{0:t-1}, y_{1:t})$ based on the application at hand. For a discussion on some of those we refer the reader to \cite{doucet2000sequential}. The authors show that the optimal choice, in the sense of minimizing the variance of the importance weights, is to select, 
\[g(x_t |x_{0:t-1}, y_{1:t}) = p(x_t|x_{t-1}, y_t) = \frac{p(x_t, y_t|x_{t-1})}{p(y_t|x_{t-1})} \propto \frac{f(y_t | x_t)q(x_t | x_{t-1})}{f(y_t|x_{t-1})}\]
Thus the particle weights from (\ref{eq: wupdate}) become $w_t^{(i)} \propto f(y_t|x_{t-1}) w_{t-1}^{(i)}$. Unfortunately, in most applications the optimal importance density might not be available in closed form and approximations or alternative specifications are required. 

A practical approach is to set the importance density function to be the transition density of the state, $q(x_t | x_{t-1})$, \cite{gordon1993novel}. While it could be beneficial to do so in some applications, if the likelihood is highly informative in relation to the posterior, there are any outliers present in the data, or the latent variable is of large dimension, the state transition density can lead to very unsatisfactory results as the state space is explored without accounting for the observations. 

Originally proposed in \cite{pitt1999filtering}, the auxiliary particle filter is a two stage algorithm that incorporates the information contained in the observed data into the importance density function. The first stage consists of sampling auxiliary variables, corresponding to the current observation. At the second stage the particles from the previous time point are resampled using those auxiliary variables. Further, the new particles are sampled from the transition density conditional on those auxiliary variables. After the sampling is concluded, the auxiliary variables are discarded. In other words as described by \cite{petris2009dynamic} we have at each iteration of the algorithm:
\begin{enumerate}
\item Draw an auxiliary variable $K$ with probability $P[K = k] \propto w_{t-1}^{(i)}\hat{p}(y_t |\xi(x_{t-1}^{(i)}))$
\item Draw $x_t^{(i)} \sim q(x_t|x_{t-1}^{(k)}) $
\end{enumerate}
The weight updates for the auxiliary particle filter become: 
\[w_{t}^{(i)} =  \frac{w_{t-1}^{(k)} p(y_t | x_t^{(i)})q(x_t^{(i)}| x_{t-1}^{(i)})}{w_{t-1}^{(k)} \hat{p}(y_t | \xi(x_{t-1}^{(i)}))q(x_t^{(i)}| x_{t-1}^{(i)})}  = \frac{p(y_t | x_t^{(i)})}{\hat{p}(y_t | \xi(x_{t-1}^{(i)}))}\]
There has been various improvements on the original APF discussed in the literature. One such work is\cite{carpenter1999improved}, who introduce a one stage auxiliary filter that is intended to reduce the variance of the original APF. It has been found that this version of the APF performs better in practice \cite{johansen2008note}. 

\section{Auxiliary Particle Approximate Bayesian Computation Filter}

The discussion of the SMC algorithms in the previous section suggests that the implementation of the algorithm relies on the knowledge of the likelihood function. In particular, in order to evaluate the weights, necessary for propagating the particles, one must evaluate the probability density function of the observed process. However, the likelihood of the $\alpha$-stable SVM is unknown and standard SMC can not be applied. To circumvent this difficulty, we consider the particle filter in ABC framework.  There are several existing papers in the literature, which incorporate ABC into SMC. \cite{jasra2012filtering} propose a SMC-ABC algorithm for state space models and apply it to a portfolio optimization problem. The authors suggest using the transition density as the importance function. As previously discussed, in situations where the data is very informative, doing so might lead to poor posterior estimates. 

Below we present our ABC auxiliary particle filter, which can be applied to any hidden Markov model, for which 
the p.d.f. is not available in closed form, but simulation from the model is possible:
\begin{algorithm}[H]
  \caption{ABC-APF}
  \begin{algorithmic}
    \State \underline{Initialization: \textit{$t = 0$}}
    \State  \ \ \ \     Sample $x_0^{(1)}, \ ... \ , x_0^{(N)}$ from $\pi_0$; set $w_0^{(i)} = \frac{1}{N}$ for $i = 1, \ ... \ , N$
     \State \underline{Update: \textit{$t \geq 1$}}
      \For  {$t=1, \ ... \ , T$} 
    \State  Set $w^{(i)}_{t-1} \propto \tilde{w}^{(i)}_{t-1} \hat{p}(y_t|\xi(x_{t-1}^{(i)})) $   
    \State   Resample N particles $x_{t-1}^{(i)}$ with respect to $w_{t-1}^{(i)}$ and set $w_{t-1}^{(i)} = \frac{1}{N} $
 \For  {$i=1, \ ... \ , N$} 
\State Sample $x_t^{(i)}$  from $p(x_t|x_{t-1} = x_{t-1}^{(i)})$
       
    \State Update the weights $\tilde{w}_t^{(i)} = \frac{K_{\epsilon}(y_t^{sim} - y_t |x_t^{(i)})q(x_t|x_{t-1})}{\hat{p}(y_t|\xi(x_{t-1}^{(i)}))p(x_t|x_{t-1})}$
   \State Normalize the weights $\tilde{w}_t^{(i)} = \tilde{w}_t^{(i)} / \sum_{j=1}^N \tilde{w}_t^{(i)}$ 
   \EndFor
    \EndFor 
  \end{algorithmic}
\end{algorithm}
The particles and weights at time t give us the desired approximation to $p(x_{0:t}|y_{1:t}) \approx \sum_{i=1}^N w_t^{(i)}\delta_{x_{0:t}}^{(i)}$. 
We include a resampling step in our algorithm, to deal with one of the inherent limitations of any SMC sampler, sample depletion. If we deal with a data set that spans large periods of time, the samples drawn from the approximate posterior distribution degenerate over time. In fact, after updating the samples for some time only a few of them have relatively large weights. This leads to a poor posterior distribution approximation, as it is represented by only a few particles. One way of dealing with this issue is to introduce a resampling step into the SMC algorithm, in order to refresh the samples. There are several different ways that this can be achieved, the simplest being the multinomial resampling. For other resampling methods and more on this issue we refer the reader to \cite{douc2005comparison}. 
In the ABC-APF algorithm as presented above, particles are resampled at every stage. 
Although this improves the estimates of the posterior distribution, it introduces extra computational cost. To avoid resampling at every single step, but to take advantage of the stability of the posterior estimates introduced by resampling, practitioners have adopted the effective sample size measure of \cite{kong1994sequential} and \cite{liu1996metropolized}. This measure was originally introduced to determine the degeneracy of the algorithm and its approximation is given by:
\[ \hat{N}_{\mbox{eff}} = \frac{1}{\sum_{i=1}^N (w_{t}^{(i)})^2} \ \ \ \ \ \forall \ t \geq 0\]
Thus, the resampling step can be carried out only if the estimated effective sample size $N_{\mbox{eff}} < N_0$ for some pre specified threshold $N_0$.

The weights in the ABC-APF depend on $\hat{p}(y_t |\xi(x_{t-1}^{(i)}))$, where $\xi: \mathcal{X} \rightarrow \mathbb{R}$. The prevalent choice in the literature has been to set  $\hat{p}(y_t |\xi(x_{t-1}^{(i)})) = f(y_t |\mathbb{E}[x_t|x_{t-1}^{(i)}])$. While, using the observation density evaluated at the expected value of the state variable is a convenient choice, \cite{johansen2008note} suggest that doing so might not lead to good estimates. In particular, the authors discuss that having an importance function with heavier tails than the target is important, because otherwise one might obtain estimates with unbounded variance. Moreover, if we assume that the observation variable follows $\alpha$-stable distribution, the density is not available in closed form, and the weights will be more difficult to update. A suitable choice that circumvents this difficulty is to take $\hat{p}(y_t |\xi(x_{t-1}^{(i)}))$  to be the $t$-distribution. 

One may note that the choice of a kernel $K_{\epsilon}(y)$ is critical to the performance of the algorithm. While, most work in the literature has used a uniform kernel, it has not been shown that doing so exhibits superior performance. In the next section we illustrate the performance of ABC-APF with $K_{\epsilon}(y) = N(0, \epsilon^2)$. 

Our proposed algorithm not only provides a solution to the SVM problem with underlying stable distribution, but it also extends the auxiliary particle filter and allows it accommodation of any hidden Markov model, in which the likelihood function is not known in a closed form. 

\section{Simulation Results}
We proceed in this section to apply our proposed algorithm to a simulated data from the model given in (\ref{eq:ssvmy}) and (\ref{eq:ssvmx}). We assume that $w_t \sim N(0, \sigma_h)$, \ $v_t \sim \mathcal{SD}(\alpha, \beta, 0, \sigma_v)$. We set the parameters of the model to the following values: $\alpha = 1.75, \ \beta = 0.1, \ \mu = -0.2, \ \phi = 0.95, \ \sigma_h = 0.6, \ \sigma_v = 0.8$. We simulate 500 observation and state data points. For the ABC-APF filter we use $N = 5000$ particles. To fully specify the algorithm we need to select $K_{\epsilon}(y_t^{sim} - y_t|h_t^{(i)})$ and $\hat{p}(y_t|\xi(h_t^{(i)}))$. We set the latter density to be the pdf of a $t$-distribution. Following our previous discussion on having a proposal distribution with heavier tails than the target distribution, we choose a $t$-distribution with 2 degrees of freedom. We consider three different versions of the $t$-distribution. We use a standard $t$-distribution, in which case we do not account for the state in our proposal, a shifted $t$-distribution pdf evaluated at $y_t - \mathbb{E}[h_t | h_t-1]$,  and a non-central $t$-distribution with location parameter $\mu + \phi h_{t-1}^{(i)}$ which is the mean of the transition density. We set $K_{\epsilon}(y_t^{sim} - y_t|h_t)$ to be a normal kernel centered at $y_t$ with standard deviation given by $\epsilon$. Figure 1 represents a plot of the true state values over plotted with the mean estimates of the latent unobserved log-volatility given by $h_t, \ t= 1, ..., 500$ from the ABC-APF filter with $\epsilon = 0.25$. We find that this value of epsilon provides good estimates from our filter.  We set $\hat{p}(y_t|\xi(h_t^{(i)}))$ to be the shifted $t$ distribution. As shown by Figure 1, our algorithm captures the general trend very well and provides estimates of the state that are close to the true values. 

\begin{figure}[H]
\centering
  \includegraphics[width = 0.81\linewidth]{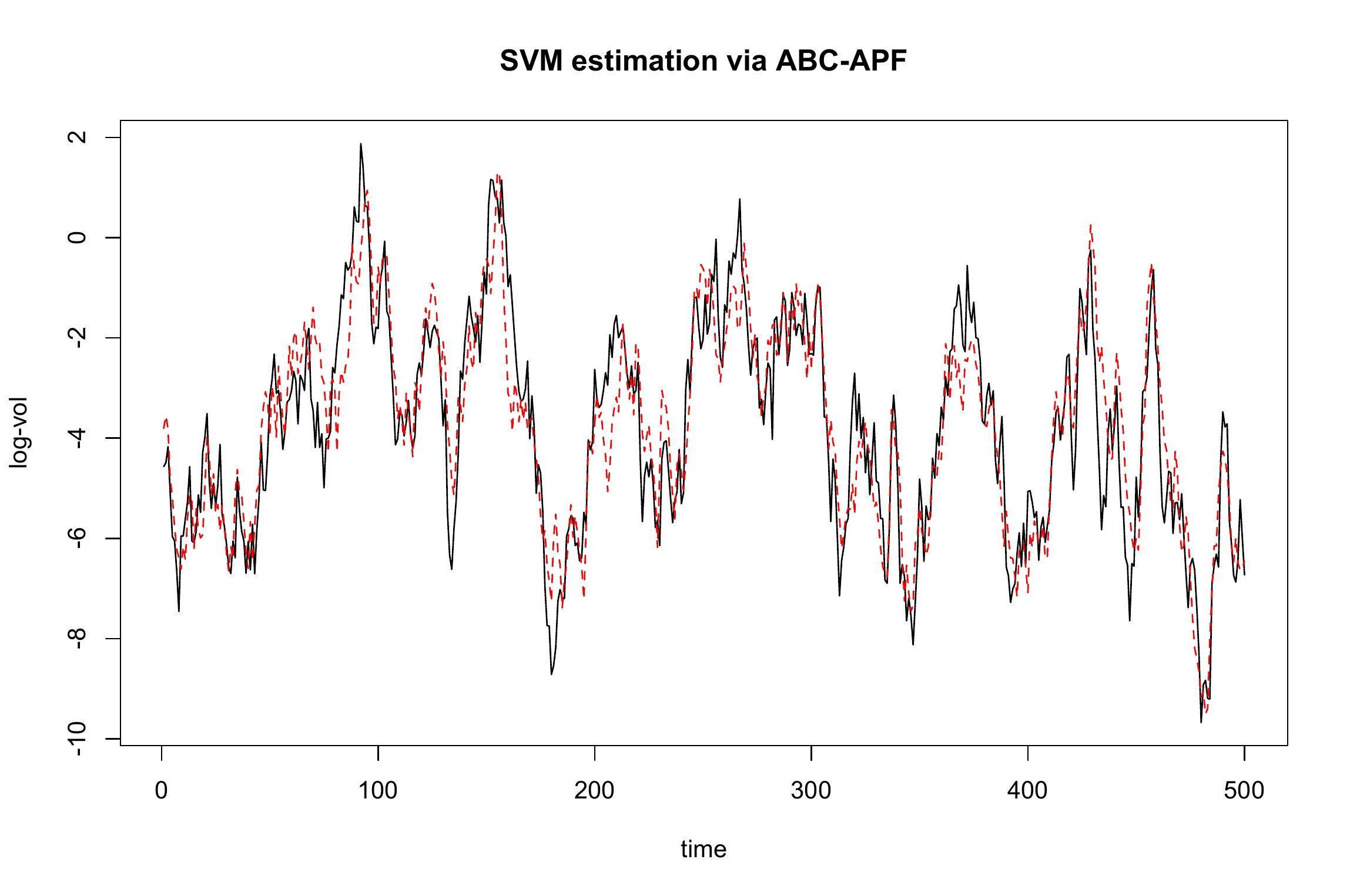}
  \caption{ABC-APF mean volatility estimate (dashed line) and the true state volatility(solid line). The filtered values presented are averaged over 100 simulations of the APF-ABC. The data is simulated from the model given by equations (\ref{eq:ssvmy}) and (\ref{eq:ssvmx}) with $w_t \sim N(0, \sigma_v)$, \ $v_t \sim \mathcal{SD}(\alpha, \beta, 0, 1)$ where  $\alpha = 1.75, \ \beta = 0.1, \ \mu = -0.2, \ \phi = 0.95, \ \sigma_h = 0.6, \ \sigma_v = 0.8$}
  \label{fig:one}
\end{figure}
To study the performance of our algorithm and the effect of the different proposal distributions, we compare the root-mean-square-error (RMSE), the absolute error (AE) and CPU times of our proposed algorithm for three alternative specifications of $\hat{p}(y_t|\xi(h_{t}))$ in Table \ref{table:one}. In addition, we compare to results obtained from the ABC-SMC method with uniform kernel as in \cite{jasra2012filtering}. The distance metric used is the absolute distance and $\epsilon$ is defined to be the percentile $P_{\epsilon}$ of shortest absolute distances between the simulated and original data. All the statistics presented in the table are based on averages from 100 runs of the Sequential Monte Carlo samplers. All algorithms are run on a 2.5GHz Intel Core i5 with 4GB RAM.

\begin{figure}[H]
\centering
  \includegraphics[width = 0.5\linewidth]{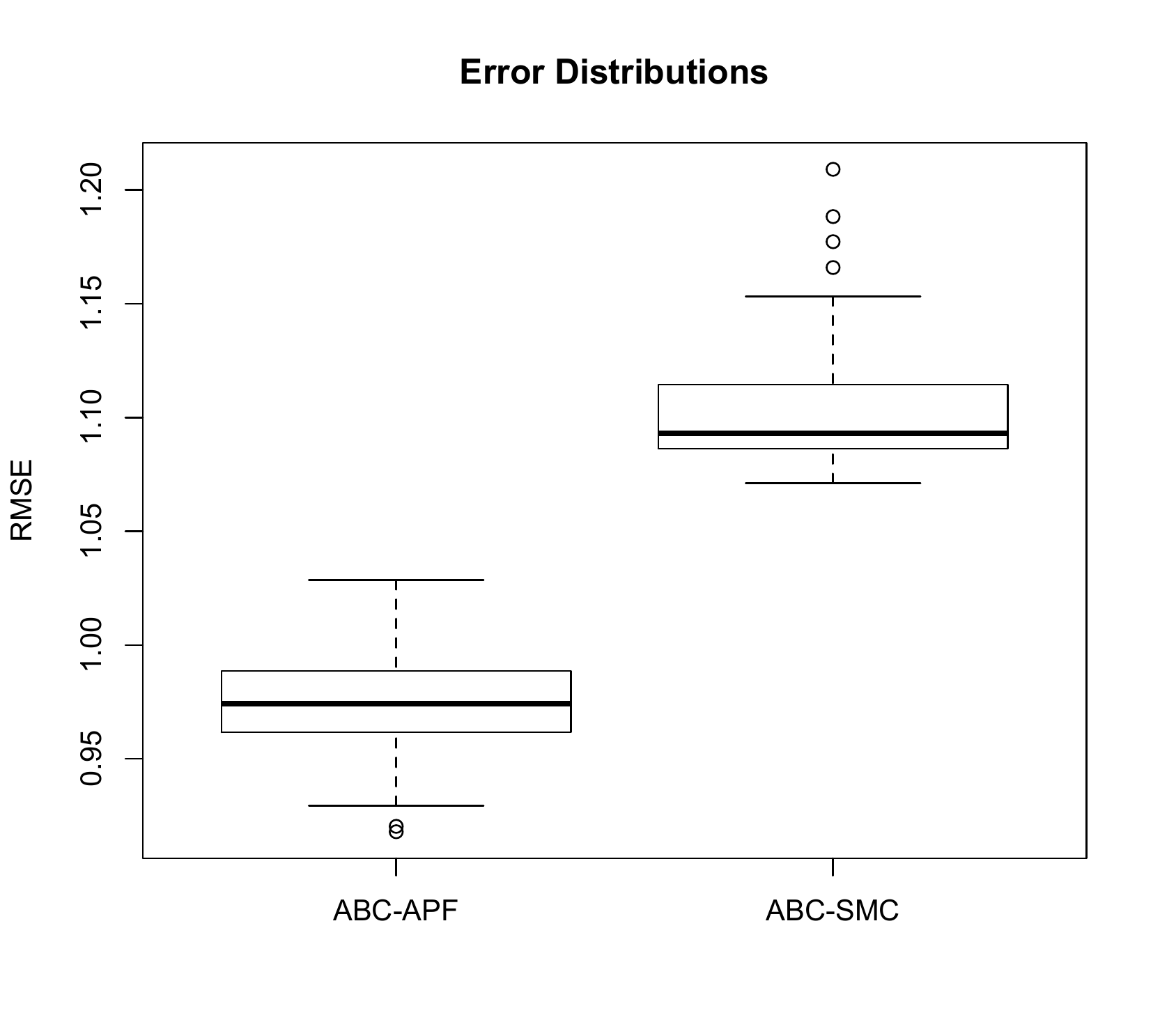}
  \caption{Box plots of the root-mean-square error for ABC-APF and ABC-SMC. For the ABC-APF, $\hat{p}(y_t|\xi(h_t^{(i)}))$ is the shifted $t$ distribution. The box plots were obtained from 100 runs of each filter.}
  \label{fig:two}
\end{figure}

In all cases, we observe relatively low errors given the range of the log-volatility, which shows us that our algorithm has performed relatively well in capturing the unobserved state. For this specification of the model, evaluating the p.d.f. of the central $t$-distribution after shifting the data by the mean of the transition density, provides desirable tradeoff between speed and accuracy of the algorithm as witnessed by the low RMSE, AE and CPU times. The non-central $t$-distribution provides good results as well, but the CPU time when using this density are much higher. This can be attributed to the computational complexity of evaluating the p.d.f. of non-central $t$-distribution.  When comparing the ABC-SMC method with uniform kernel one notices that the RMSE and AE error are slightly higher than the APF-ABC for the shifted and non-central $t$-distribution cases. We attribute this behavior to the fact that in the ABC-SMC algorithm the proposed particles at each step do not account for the current data point. To further study the differences between ABC-APF and ABC-SMC, we present in Figure \ref{fig:two} box plots of the RMSE based on the shifted $t$-distribution for the ABC-APF. 

Figure \ref{fig:two} illustrates the advantage of our proposed filter, with the maximum error of ABC-APF less than the minimum possible error for ABC-SMC. This indicates indicates that on every run, our filter provides lower root-mean-squared error. In fact, as \cite{pitt1999filtering} suggest, for the stochastic volatility model, a specific case of the more general class of hidden Markov models, the conditional likelihood of the data is not very sensitive to the values of the state. In other words, for other hidden Markov models, the expected performance of ABC-APF is even greater than that achieved based on the stochastic volatility model discussed here. It is worth nothing that the decreases in accuracy for the ABC-SMC with uniform kernel are more drastic. This is expected since we set epsilon to be the percentile $P_{\epsilon}$ of shortest absolute distances. 
\begin {table}[H]%
\begin{center}
\begin {tabular}{clcccc}
\toprule%
& &  & &\\
$\epsilon$& &0.25&0.5 &0.75&1.5  \\\toprule%
RMSE & central $t$ & 1.051& 1.044 &  1.049 & 1.099 \\
 & shifted $t$ & 0.984 & 0.981 &  0.988 & 1.067 \\
& non-central $t$ & 1.032 & 1.032 & 1.059& 1.259 \\
& ABC-SMC Uniform kernel& 1.102 & 1.248 & 1.634 & -\\\midrule
AE & central $t$ & 0.817 & 0.813 & 0.82 & 0.87\\
 & shifted $t$ & 0.755 & 0.754 &  0.763 & 0.829 \\
& non-central $t$ & 0.794  & 0.797 & 0.816 & 0.958  \\
& ABC-SMC Uniform Kernel& 0.863 & 0.99 & 1.329 & -  \\\midrule
CPU time & central $t$ &2.35&2.34&2.35&2.33 \\
 & shifted $t$ & 2.81 & 2.80&2.79&2.80 \\
& non-central $t$ & 17.36 & 17.56 & 17.4 & 17.28  \\
& ABC-SMC Uniform Kernel & 2.70 & 2.67 & 2.5 & - \\\bottomrule
\end {tabular}
 \caption{APF-ABC with different choices of $\hat{p}(y_t|\hat{h}_{t})$ and Gaussian kernel, and ABC-SMC with uniform kernel as in \cite{jasra2012filtering}. All values are averaged over 100 filter runs. All CPU times are given as average time in seconds per run. CPU specifications are: 2.5GHz Intel Core i5 with 4GB RAM}\label {aggiungi}\centering%
\label{table:one}
\end{center}
\end {table}
%$p(y_t|x_t^{i})$ in the weight update step by 
%\[ \pi(y_{sim}, y|x_t^{i}) = I_{\Omega_\epsilon}   \]
%where $\Omega_\epsilon = {y_{sim}: $

In Table \ref{table:two} we explore the sensitivity of ABC-APF to various parameter values. The first column contains five different sets of values for $\alpha$ and $\beta$. Recall that a random variable from $\alpha$-stable distribution with $\alpha = 2$ is normally distributed. We set $\mu = 0$ and $\phi = 0.9$ in equations (\ref{eq:ssvmy}) and (\ref{eq:ssvmx}). We select low ($\sigma_h = 0.2$, $\sigma_y = 0.1$) and high values ($\sigma_h = 1$, $\sigma_y = 1$).  Our goal is to study how the errors compare for similar ranges of the simulated data. We calculate the average RMSE and AE over 100 filter simulations for each set of parameter values. We choose the shifted $t$-distribution for our proposal distribution for the ABC-APF. According to table \ref{table:two} as the value of $\alpha$ moves away from 2, the performance of the filter deteriorates. This is expected as random variables generated from a stable distribution with values of $\alpha$ close to 0 have distributions with very heavy tails, which makes state filtering more difficult. We further note that as the standard deviation of the noise terms increases so do the root mean square and absolute errors. This is a direct result of the wider range of true values generated from the model due to the higher variability in the noise. \\

\begin{table}[H]
  \centering
  \begin{tabular}{c|c|c|c|c}
   
    & \multicolumn{2}{|c}{$\sigma_h = 0.2$, $\sigma_y = 0.1$} & \multicolumn{2}{|c}{$\sigma_h = 1$, $\sigma_y = 1$} \\
    \hline
    \multicolumn{1}{c|}{} & RMSE & AE & RMSE & AE \\
    \hline
    \multicolumn{1}{r|}{$\alpha = 2, \beta = \mbox{NA}$} &$0.63$ &$0.523$&  $1.01$ & $0.791$\\
    \multicolumn{1}{r|}{$\alpha = 1.9, \beta = 0.9$} & $0.723$ &$0.564$& $1.053$ & $0.831$\\
    \multicolumn{1}{r|}{$\alpha = 1.2, \beta = 0.3$} & $0.806$ &$0.65$& $1.232$ & $0.981$ \\
    \multicolumn{1}{r|}{$\alpha = 0.8, \beta = -0.2$} & $0.819$ &$0.656$& $1.561$ & $1.252$ \\
    \multicolumn{1}{r|}{$\alpha = 0.1, \beta = -0.8$} & $0.885$ &$0.64$& $1.858$ & $1.357$ \\
    \hline
  \end{tabular}
  \caption{APF-ABC with different parameter values. All values are averaged over 100 filter runs. Root mean squared error (RMSE) and Absolute Error (AE). Note $\alpha =2$ is equivalent to a normal distribution}
\label{table:two}

\end{table}

\begin{figure}[H]
 \includegraphics[width = 0.52\linewidth]{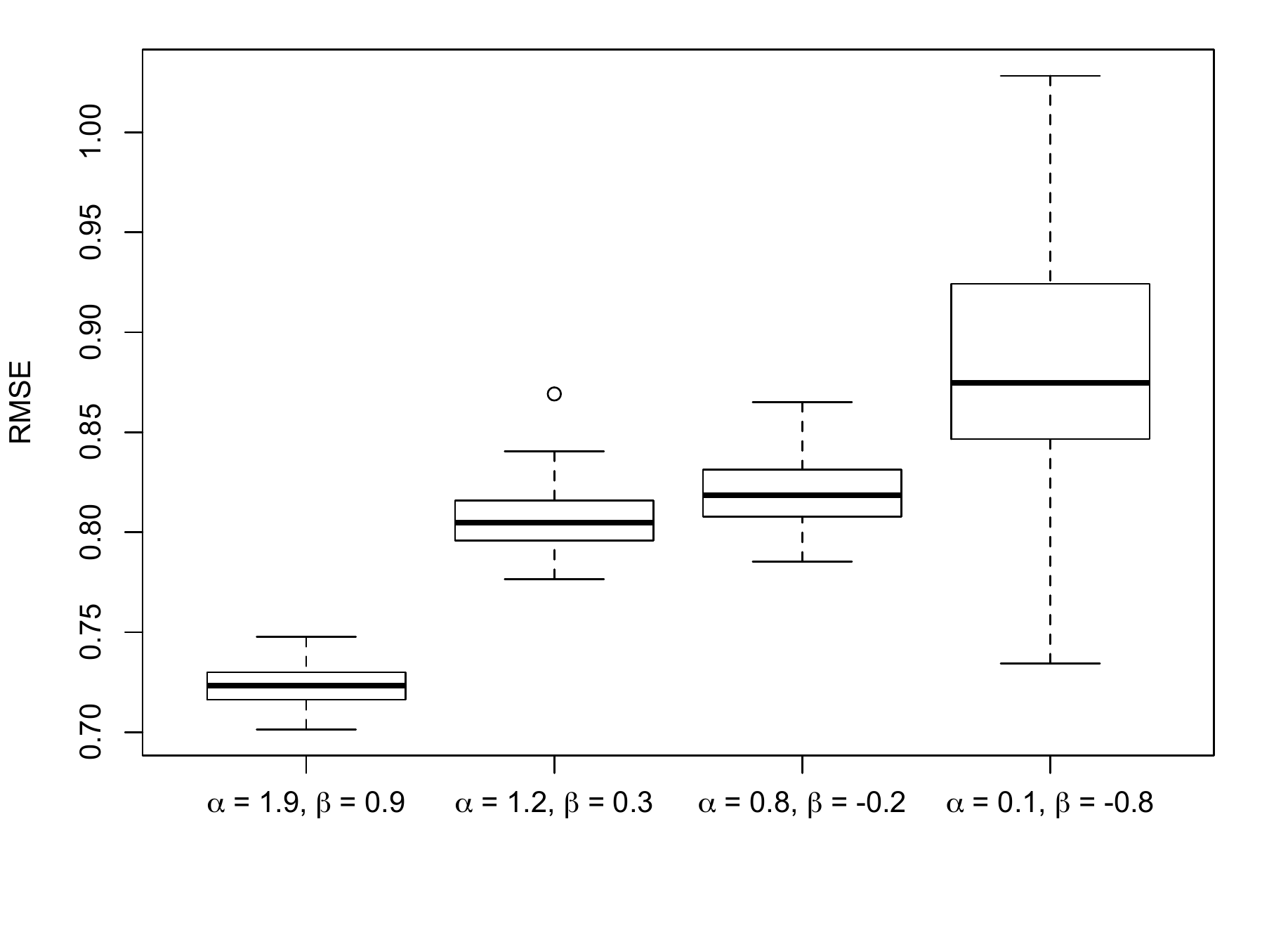}%
  \includegraphics[width = 0.52\linewidth]{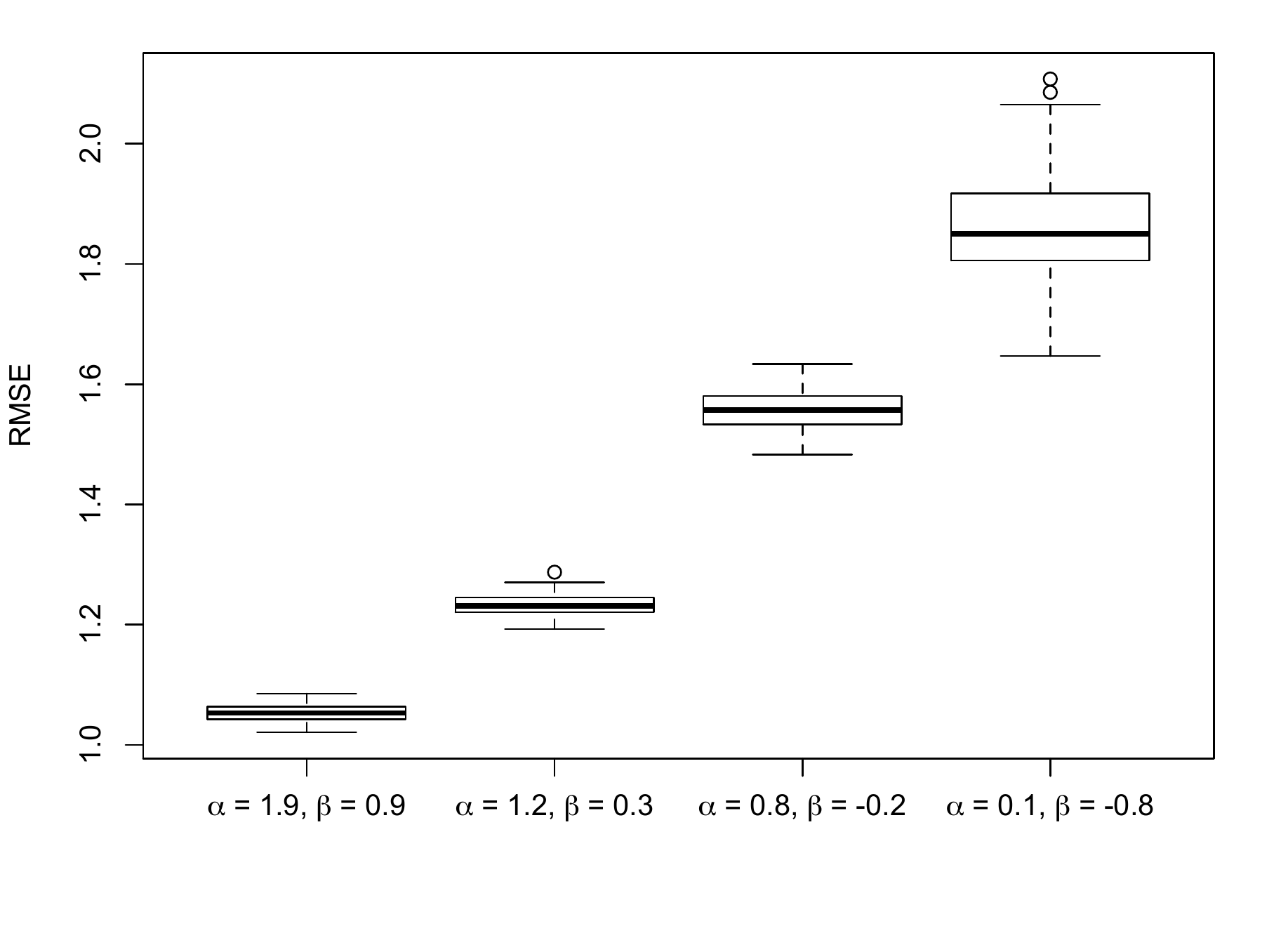}
  \caption{Box plots for the root mean squared error for different values of $\alpha$ and $\beta$. All box plots are created based on 100 runs from the ABC-APF. The simulation assumes the values $\sigma_h = 0.2$, $\sigma_y = 0.1$ (left panel) and $\sigma_h = 1$, $\sigma_y = 1$ (right panel). The values $\phi = 0.9$, $\mu = 0$ are used in both figures.}
  \label{fig:three}
\end{figure}

\begin{figure}[H]
 \includegraphics[width = 0.6\linewidth]{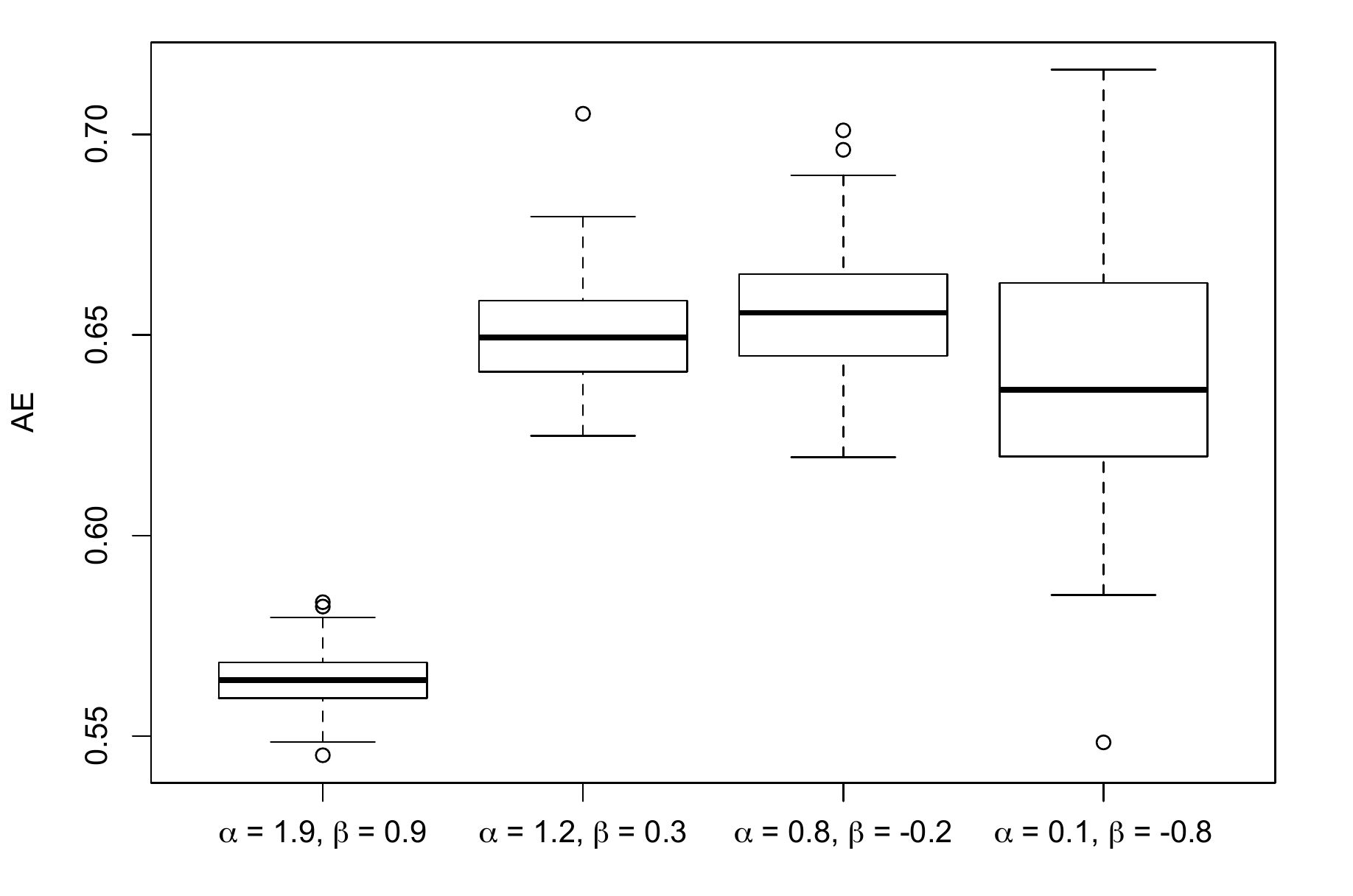}%
  \includegraphics[width = 0.46\linewidth]{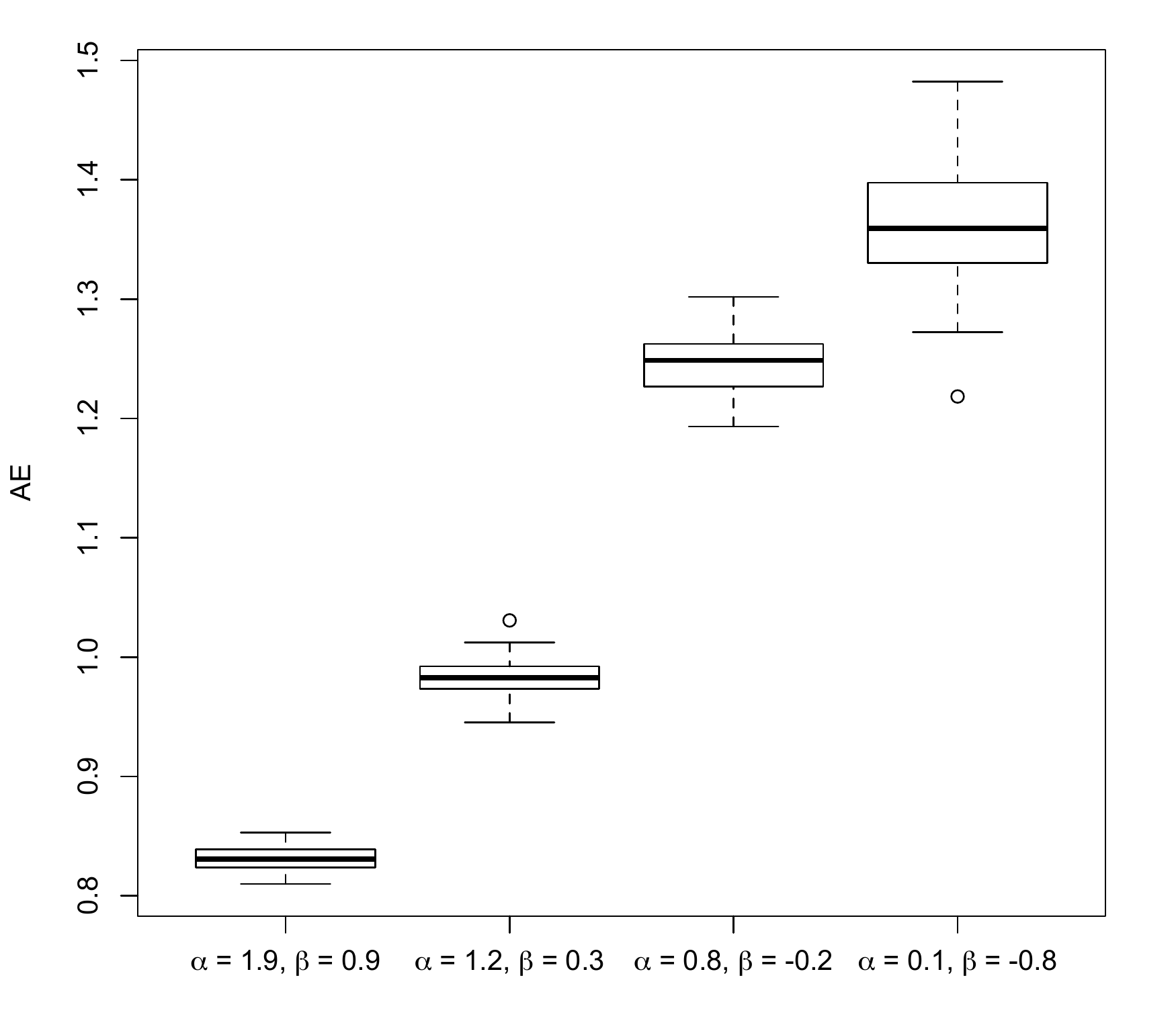}
  \caption{Box plots for the absolute error for different values of $\alpha$ and $\beta$. All box plots are created based on 100 runs from the ABC-APF. The simulation assumes the values $\sigma_h = 0.2$, $\sigma_y = 0.1$ (left panel) and $\sigma_h = 1$, $\sigma_y = 1$ (right panel). The values $\phi = 0.9$, $\mu = 0$ are used in both figures.}
  \label{fig:four}
\end{figure}

Figures \ref{fig:three} and \ref{fig:four} give an overview of the empirical distribution of the root mean square and absolute errors respectively. The box plots further confirm that decreasing the value of $\alpha$ increases both errors, regardless of the standard deviation of the underlying noise terms. In the left panel of figure \ref{fig:three} the median value for the absolute error in the scenario when $\alpha = 0.1$, $\beta = -0.8$ is the second lowest. However, we note that the absolute error distributions have a much wider range.

\section{Conclusion}
In this paper we have considered a class of stochastic volatility models, which allows for skewness and heavy tails in the returns. Doing so aligns better with what empiricists have observed in stock price data. In our particular study we have assumed that the data follows $\alpha$-stable distribution. As the likelihood of the model is not available in closed form, standard SMC algorithms do not apply. We present a new approach which enables us to take advantage of the power of the auxiliary filter and at the same time allows us to use it in settings, where it was not possible before. To accomplish our goal we take advantage of the newly developed and upcoming paradigm of approximate Bayesian computation. To illustrate our proposed methodology and its application to SVM we provide a simulation study, in which we obtain encouraging results. Our filter is able to correctly capture the trend of the underlying latent log-volatility, the true value for which is available in our study. We also present an error analysis, which confirms the usefulness of our algorithm. 

The ABC auxiliary particle filter that we propose here can be applied to any HMM, for which the likelihood function is either not available in closed form or difficult to compute. Our proposed filter is very fast and can be easily extended to various situations that require use of particle filters. Developing this auxiliary particle filter provides more flexibility in model assumptions, as one no longer has to limit themselves to models for which the distribution requires the likelihood to be known. \\

\bibliographystyle{plain}
\bibliography{StochasticVolatilityFilteringwithIntractableLikelihoods}
\end{document}